*Article*

# Adaptative Perturbation Patterns: Realistic Adversarial Learning for Robust Intrusion Detection

**João Vitorino** [0000-0002-4968-3653] *****, **Nuno Oliveira** [0000-0002-5030-7751] **and Isabel Praça** [0000-0002-2519-9859] *****

Research Group on Intelligent Engineering and Computing for Advanced Innovation and Development (GECAD), School of Engineering, Polytechnic of Porto (ISEP/IPP), 4249-015 Porto, Portugal; nunal@isep.ipp.pt
***** Correspondence: jpmvo@isep.ipp.pt (J.V.); icp@isep.ipp.pt (I.P.)

**Abstract:** Adversarial attacks pose a major threat to machine learning and to the systems that rely on it. In the cybersecurity domain, adversarial cyber-attack examples capable of evading detection are especially concerning. Nonetheless, an example generated for a domain with tabular data must be realistic within that domain. This work establishes the fundamental constraint levels required to achieve realism and introduces the Adaptative Perturbation Pattern Method (A2PM) to fulfill these constraints in a gray-box setting. A2PM relies on pattern sequences that are independently adapted to the characteristics of each class to create valid and coherent data perturbations. The proposed method was evaluated in a cybersecurity case study with two scenarios: Enterprise and Internet of Things (IoT) networks. Multilayer Perceptron (MLP) and Random Forest (RF) classifiers were created with regular and adversarial training, using the CIC-IDS2017 and IoT-23 datasets. In each scenario, targeted and untargeted attacks were performed against the classifiers, and the generated examples were compared with the original network traffic flows to assess their realism. The obtained results demonstrate that A2PM provides a scalable generation of realistic adversarial examples, which can be advantageous for both adversarial training and attacks.

## 1. Introduction

Machine learning is transforming the way modern organizations operate. It can be used to automate and improve various business processes, ranging from the recognition of patterns and correlations to complex regression and classification tasks. However, adversarial attacks pose a major threat to machine learning models and to the systems that rely on them. A model can be deceived into predicting incorrect results by slightly modifying original data, which creates an adversarial example. This is especially concerning for the cybersecurity domain because adversarial cyber-attack examples capable of evading detection can cause significant damage to an organization [1], [2].

Depending on the utilized method, the data perturbations that result in an adversarial example can be created in one of three settings: black-, gray- and white-box. The first solely queries a model's predictions, whereas the second may also require knowledge of its structure or the utilized feature set, and the latter needs full access to its internal parameters. Even though machine learning is inherently susceptible to these examples, a model's robustness can be improved by various defense strategies. A standard approach is performing adversarial training, a process where the training data is augmented with examples generated by one or more attack methods [3], [4].

Nonetheless, a method can only be applied to a given domain if the examples it generates are realistic within that domain. In cybersecurity, a domain with tabular data, if an adversarial example does not resemble real network traffic, a Network-based Intrusion

Detection System (NIDS) will never actually encounter it because it cannot be transmitted through a computer network. Furthermore, if an example can be transmitted but is incompatible with its intended malicious purpose, evading detection will be futile because no damage can be caused. Consequently, training machine learning models with unrealistic cyber-attack examples only deteriorates their generalization to real computer networks and attack scenarios. Therefore, the generation of realistic adversarial examples for domains with tabular data is a pertinent research topic.

This work addressed the challenge of generating realistic examples, with a focus on network-based intrusion detection. The main contributions are the establishment of the fundamental constraint levels required to achieve realism and the introduction of the Adaptative Perturbation Pattern Method (A2PM) to fulfil these constraints in a gray-box setting. The capabilities of the proposed method were evaluated in a cybersecurity case study with two scenarios: Enterprise and Internet of Things (IoT) networks. It generated adversarial network traffic flows for multi-class classification by creating data perturbations in the original flows of the CIC-IDS2017 and IoT-23 datasets.

Due to the noticeably different internal mechanics of an Artificial Neural Network (ANN) and a tree-based algorithm, the study analyzed the susceptibility of both types of models to the examples created by A2PM. A total of four Multilayer Perceptron (MLP) and four Random Forest (RF) classifiers were created with regular and adversarial training, and both targeted and untargeted attacks were performed against them. To provide a thorough analysis, example realism and time consumption were assessed by comparing the generated examples with the corresponding original flows and recording the time required for each A2PM iteration.

The present article is organized into multiple sections. Section 2 defines the fundamental constraint levels and provides a survey of previous work on adversarial examples. Section 3 describes the proposed method and the key concepts it relies on. Section 4 presents the case study and an analysis of the obtained results. Finally, Section 5 addresses the main conclusions and future work.

## 2. Related Work

In recent years, adversarial examples have drawn attention from a research perspective. However, since the focus has been the image classification domain, the generation of realistic examples for domains with tabular data remains a relatively unexplored topic. The common adversarial approach is to exploit the internal gradients of an ANN in a white-box setting, creating unconstrained data perturbations [5]–[7]. Consequently, most state-of-the-art methods do not support other types of machine learning models nor other settings, which severely limits their applicability to other domains. This is a pertinent aspect of the cybersecurity domain, where white-box is a highly unlikely setting. Considering that a NIDS is developed in a secure context, an attacker will commonly face a black-box setting, or occasionally gray-box [8], [9].

The applicability of a method for adversarial training is significantly impacted by the models it can attack. Despite an adversarially robust generalization still being a challenge, significant progress has been made in ANN robustness research [10]–[14]. However, various other types of algorithms can be used for a classification task. This is the case of network-based intrusion detection, where tree-based algorithms, such as RF, are remarkably well-established [15], [16]. They can achieve a reliable performance on regular network traffic, but their susceptibility to adversarial examples must not be disregarded. Hence, these algorithms can benefit from adversarial training and several defense strategies have been developed to intrinsically improve their robustness [17]–[20].

In addition to the setting and the supported models, the realism of the examples generated by a method must also be considered. Martins et al. [21] performed a systematic review of recent developments in adversarial attacks and defenses for cybersecurity, and observed that none of the reviewed articles evaluated the applicability of the generated examples to a real intrusion detection scenario. Therefore, it is imperative to establish the fundamental constraints an example must comply with to be applicable to a real scenario on a domain with tabular data. We define two constraint levels:
1. Domain constraints – Specify the inherent structure of a domain;
2. Class-specific constraints – Specify the characteristics of a class.

To be valid on a given domain, an example can solely reach the first level. Nonetheless, full realism is only achieved when it is also coherent with the distinct characteristics of its class, reaching the second. In a real scenario, each level will contain concrete constraints for the utilized data features. These can be divided into two types:
- Intra-feature constraints – Restrict the value of a single feature;
- Inter-feature constraints – Restrict the values of one or more features according to the values present in other features.

In a real computer network, an example must fulfil the domain constraints of the utilized communication protocols and the class-specific constraints of each type of cyber-attack. Apruzzese et al. [8] proposed a taxonomy to evaluate the feasibility of an adversarial attack against a NIDS, based on access to the training data, knowledge of the model and feature set, reverse engineering and manipulation capabilities. It can provide valuable guidance to establish the concrete constraints of each level for a specific system.

Even though some methods attempt to fulfil a few constraints, many exhibit a clear lack of realism. Table 1 summarizes the characteristics of the most relevant methods of the current literature, including the constraint levels they attempt to address. The keyword 'CP' corresponds to any model that can output class probabilities for each data sample, instead of a single class label.

**Table 1.** Summary of relevant methods and addressed constraint levels.

| Method | Setting | Supported Models | Domain Constraints | Class-specific Constraints |
|---|---|---|---|---|
| FGSM [3] | White-box | ANN | ✘ | ✘ |
| C&W [22] | White-box | ANN | ✘ | ✘ |
| DeepFool [23] | White-box | ANN | ✘ | ✘ |
| Houdini [24] | White-box | ANN | ✘ | ✘ |
| StrAttack [25] | White-box | ANN | ✘ | ✘ |
| ZOO [26] | White-box | ANN | ✘ | ✘ |
| JSMA [27] | White-box | ANN | ✓ | ✘ |
| Polymorphic [28] | Gray-box | ANN | ✘ | ✓ |
| Reconstruction [29] | Gray-box | ANN | ✘ | ✘ |
| OnePixel [30] | Black-box | CP | ✓ | ✘ |
| RL-S2V [31] | Black-box | CP | ✘ | ✘ |
| BMI-FGSM [32] | Black-box | Any | ✘ | ✘ |
| GAN [33] | Black-box | Any | ✘ | ✘ |
| WGAN [34] | Black-box | Any | ✘ | ✘ |
| Boundary [35] | Black-box | Any | ✘ | ✘ |
| Query-Efficient [36] | Black-box | Any | ✘ | ✘ |

Regarding the Polymorphic attack [28], it addresses the preservation of original class characteristics. Chauhan et al. developed it for the cybersecurity domain, to generate examples compatible with a cyber-attack's purpose. The authors start by applying a feature selection algorithm to obtain the most relevant features for the distinction between benign network traffic and each cyber-attack. Then, the values of the remaining features, which are considered irrelevant for the classification, are perturbed by a Wasserstein Generative Adversarial Network (WGAN) [34]. On the condition that there are no class-specific constraints for the remaining features, this approach could improve the coherence of an example with its class. Nonetheless, the unconstrained perturbations created by WGAN disregard the domain structure, which inevitably leads to invalid examples.

On the other hand, both the Jacobian-based Saliency Map Attack (JSMA) [27] and the OnePixel attack [30] could potentially preserve a domain structure. The former was developed to minimize the number of modified pixels in an image, requiring full access to the internal gradients of an ANN, whereas the latter only modifies a single pixel, based on the class probabilities predicted by a model. These methods perturb the most appropriate features without affecting the remaining features, which could be beneficial for tabular data. However, neither validity nor coherence can be ensured because they do not account for any constraint when creating the perturbations.

To the best of our knowledge, no previous work has introduced a method capable of complying with the fundamental constraints of domains with tabular data, which hinders the development of realistic attack and defense strategies. This is the gap in the current literature addressed by the proposed method.

## 3. Proposed Method

A2PM was developed with the objective of generating adversarial examples that fulfil both domain and class-specific constraints. It benefits from a modular architecture to assign an independent sequence of adaptative perturbation patterns to each class, which analyze specific feature subsets to create valid and coherent data perturbations. Even though it can be applied in a black-box setting, the most realistic examples are obtained in gray-box, with only knowledge of the feature set. To fully adjust it to a domain, A2PM only requires a simple base configuration for the creation of a pattern sequence. Afterwards, realistic examples can be generated from original data to perform adversarial training or to directly attack a classifier in an iterative process (Figure 1).

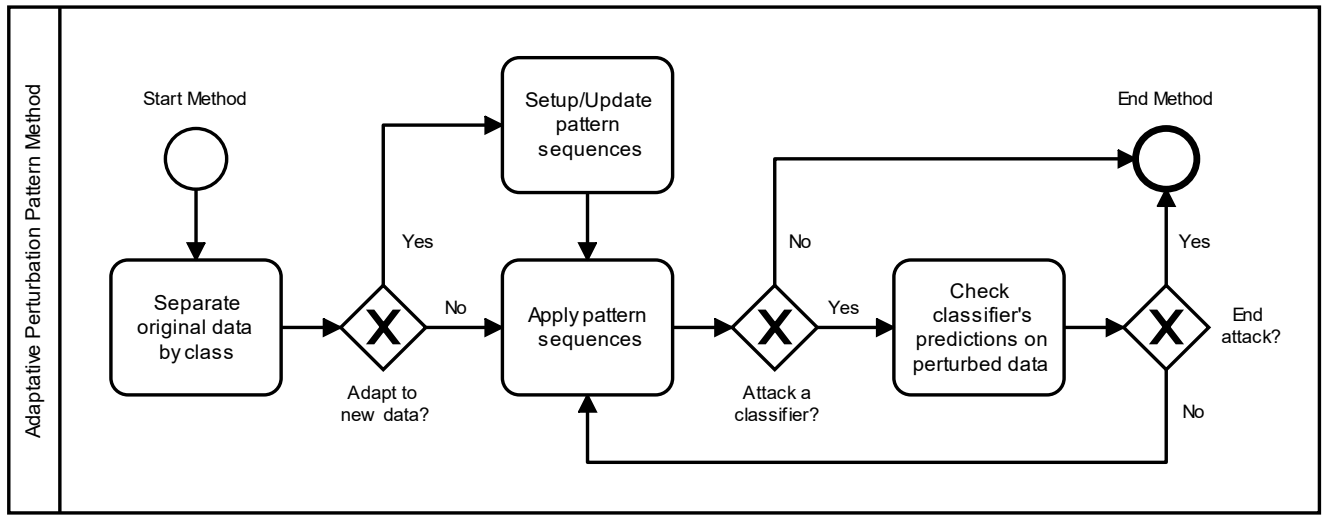

**Figure 1.** Adaptative Perturbation Pattern Method (business process model and notation).

The generated examples can be untargeted, to cause any misclassification, or targeted, seeking to reach a specific class. New data perturbations could be generated indefinitely, but it would be computationally expensive. Hence, early stopping is employed to end the attack when the latest iterations could not cause any further misclassifications. Besides static scenarios where the full data is available, the proposed method is also suitable for scenarios where it is provided over time. After the pattern sequences are created for an initial batch of data, these can be incrementally adapted to the characteristics of subsequent batches. If novel classes are provided, the base configuration is used to autonomously create their respective patterns.

The performed feature analysis relies on two key concepts: value intervals and value combinations. The following subsections detail the perturbation patterns built upon these concepts, as well as the advantages of applying them in sequential order.

*3.1. Interval Pattern*

To perturb uncorrelated numerical variables, the main aspect to be considered is the interval of values each one can assume. This is an intra-feature constraint that can be fulfilled by enforcing minimum and maximum values.

The Interval pattern encapsulates a mechanism that records the valid intervals to create perturbations tailored to the characteristics of each feature (Figure 2). It has a configurable 'probability to be applied', in the $(0, 1]$ interval, which is used to randomly determine if an individual feature will be perturbed or not. Additionally, it is also possible to specify only integer perturbations for specific features.

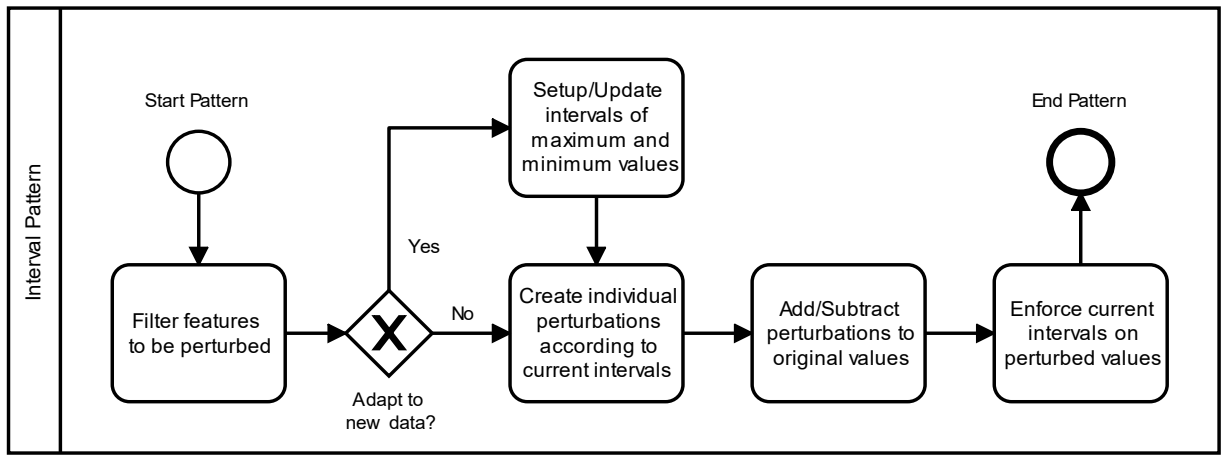

**Figure 2.** Interval pattern (business process model and notation).

Instead of a static interval, moving intervals can be utilized after the first batch to enable an incremental adaptation to new data, according to a configured momentum. For a given feature and a momentum $k \in [0, 1]$, the updated minimum $m_i$ and maximum $M_i$ of a batch $i$ are mathematically defined as:

$$m_i = m_{i-1} * k + \min(x_i) * (1 - k) \quad (1)$$

$$M_i = M_{i-1} * k + \max(x_i) * (1 - k) \quad (2)$$

where $\min(x_i)$ and $\max(x_i)$ are the actual minimum and maximum values of the samples $x_i$ of batch $i$.

Each perturbation is computed according to a randomly generated number and is affected by the current interval, which can be either static or moving. The random number $\varepsilon \in (0, 1]$ acts as a ratio to scale the interval. To restrict its possible values, it is generated within the standard range of $[0.1, 0.3]$, although other ranges can be configured. For a given feature, a perturbation $P_i$ of a batch $i$ can be represented as:

$$P_i = (M_i - m_i) * \varepsilon \quad (3)$$

After a perturbation is created, it is randomly added or subtracted to the original value. Exceptionally, if the original value is less or equal to the current minimum, it is always increased, and vice-versa. The resulting value is capped at the current interval to ensure it remains within the valid minimum and maximum values of that feature.

*3.2. Combination Pattern*

Regarding uncorrelated categorical variables, enforcing their limited set of qualitative values is the main intra-feature constraint. Therefore, the interval approach cannot be replicated even if they are encoded to a numerical form, and a straightforward solution can be recording each value a feature can assume. Nonetheless, the most pertinent aspect of perturbing tabular data is the correlation between multiple variables. Since the value present in a variable may influence the values used for other variables, there can be several inter-feature constraints. To improve beyond the previous solution and fulfil both types of constraints, several features can be combined into a single common record.

The Combination pattern records the valid combinations to perform a simultaneous and coherent perturbation of multiple features (Figure 3). It can be configured with locked features, whose values are used to find combinations for other features without being modified. Due to the simultaneous perturbations, its 'probability to be applied', in the (0, 1] interval, can affect several features.

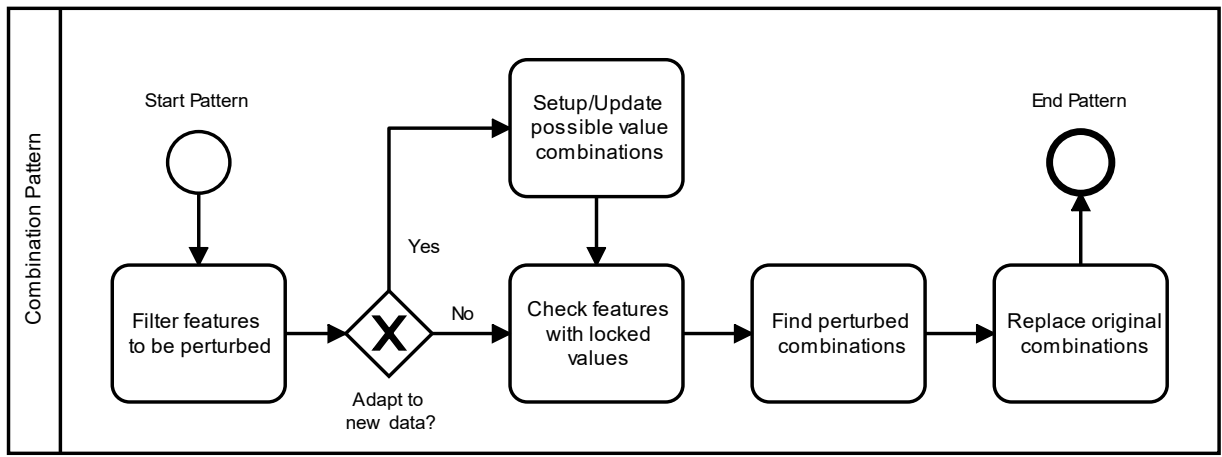

**Figure 3.** Combination pattern (business process model and notation).

Besides the initially recorded combinations, new data can provide additional possibilities. These can be merged with the previous or used as gradual updates. For a given feature and a momentum $k \in [0, 1]$, the number of updated combinations $C_i$ of a batch $i$ is mathematically expressed as:

$$C_i = C_{i-1} * k + \text{unique}(x_i) \qquad (4)$$

where $\text{unique}(x_i)$ is the number of unique combinations of the samples $x_i$ of batch $i$.

Each perturbation created by this pattern consists of a combination randomly selected from the current possibilities, considering the locked features. It directly replaces the original values, ensuring that the features remain coherent.

*3.3. Pattern Sequences*

Domains with diverse constraints may require an aggregation of several Interval and Combination patterns, which can be performed by pattern sequences. Furthermore, the main advantage of applying multiple patterns in a sequential order is that it enables the fulfilment of countless inter-feature constraints of greater complexity. It is pertinent to note that all patterns in a sequence are independently adapted to the original data, to prevent any bias when recording its characteristics. Afterwards, the sequential order is enforced to create cumulative perturbations on that data.

To exemplify the benefits of using these sequences, a small but relatively complex domain will be established. It contains three nominal features, F0, F1 and F2, and two integer features, F3 and F4. For an adversarial example to be realistic within this domain, it must comply with the following constraints:
- F0 must always keep its original value;
- F1 and F4 can be modified but must have class-specific values;
- F2 and F3 can be modified but must have class-specific values, which are influenced by F0 and F1.

The base configuration corresponding to these constraints specifies the feature subsets that each pattern will analyze and perturb:
1. Combination pattern – Modify {F1};
2. Combination pattern – Modify {F2, F3}, Lock {F0, F1};
3. Interval pattern – Modify {F3, F4}, Integer {F3, F4}.

A2PM will then assign each class to its own pattern sequence. For this example, the 'probability to be applied' will be 1.0 for all patterns, to demonstrate all three cumulative perturbations (Figure 4). The first perturbation created for each class is replacing F1 with another valid qualitative value, from 'B' to 'C'. Then, without modifying the original F0 nor the new F1, a valid combination is found for F0, F1, F2 and F3. Since the original F2 and F3 were only suitable for 'A' and 'B', new values are found to match 'A' and 'C'. Finally, the integer features F3 and F4 are perturbed according to their valid intervals. Regarding F3, to ensure it remains coherent with F0 and F1, the perturbation is created on the value of the new combination.

|  | F0 | F1 | F2 | F3 | F4 |
|---|---|---|---|---|---|
| Original | A | B | H (AB) | 21 (AB) | 47 |
| Pattern 1 (Combination) | A | C | H | 21 | 47 |
| Pattern 2 (Combination) | A | C | T (AC) | 85 (AC) | 47 |
| Pattern 3 (Interval) | A | C | T | 83 | 49 |

▨ Locked Features   ▢ Modified Features

**Figure 4.** Exemplification of a perturbation pattern sequence.

## 4. Experimental Evaluation

A case study was conducted to evaluate the capabilities of the proposed method, as well as its suitability for multi-class classification on the cybersecurity domain. Assessments of example realism and time consumption were performed by comparing the examples generated by A2PM with the original data and recording the time required for each iteration. To thoroughly analyze example realism, the assessment included examples generated by the potential alternatives of the current literature: JSMA and OnePixel.

Since the internal mechanics of an ANN and a tree-based algorithm are noticeably different, the susceptibility of both types of models to A2PM was analyzed by performing targeted and untargeted attacks against MLP and RF classifiers. Two scenarios were considered: Enterprise and IoT networks. For these scenarios, adversarial network traffic flows were generated using the original flows of the CIC-IDS2017 and the IoT-23 datasets, respectively. In addition to evaluating the robustness of models created with regular training, the effects of performing adversarial training with A2PM were also analyzed.

The study was conducted on relatively common hardware: a machine with 16 gigabytes of random-access memory, an 8-core central processing unit, and a 6-gigabyte graphics processing unit. The implementation relied on the Python 3 programming language and several libraries: *Numpy* and *Pandas* for data preprocessing and manipulation, *Tensorflow* for the MLP models, *Scikit-learn* for the RF models, and *Adversarial-Robustness-Toolbox* for the alternative methods. The following subsections describe the most relevant aspects of the case study and present an analysis of the obtained results.

*4.1. Datasets and Data Preprocessing*

Both CIC-IDS2017 and IoT-23 are public datasets that contain multiple labeled captures of benign and malicious network flows. The recorded data is extremely valuable for intrusion detection because it includes various types of common cyber-attacks and manifests real network traffic patterns.

CIC-IDS2017 [37] consists of 7 captures of cyber-attacks performed on a standard Enterprise computer network with 25 interacting users. It includes Denial-of-Service and Brute-Force attacks, which were recorded in July 2017 and are available at the Canadian Institute for Cybersecurity. In contrast, IoT-23 [38] is directed at the emerging IoT networks, with wireless communications between interconnected devices. It contains network traffic created by malware attacks targeting IoT devices between 2018 and 2019, divided into 23 captures and available at the Stratosphere Research Laboratory.

From each dataset, two captures were selected and merged, to be utilized for the corresponding scenario. Table 2 provides an overview of their characteristics, including the class proportions and the label of each class, either 'Benign' or a specific type of cyber-attack. The 'PartOfAHorizontalPortScan' label was shortened to 'POAHPS'.

**Table 2.** Main characteristics of utilized datasets.

| Scenario | Dataset (Captures) | Total Samples | Class Samples | Class Label |
|---|---|---|---|---|
| Enterprise Network | CIC-IDS2017 (Tuesday and Wednesday) | 1,138,612 | 873,066 | Benign |
| | | | 230,124 | Hulk |
| | | | 10,293 | GoldenEye |
| | | | 7,926 | FTP-Patator |
| | | | 5,897 | SSH-Patator |
| | | | 5,796 | Slowloris |
| | | | 5,499 | Slowhttptest |
| | | | 11 | Heartbleed |
| IoT Network | IoT-23 (1-1 and 34-1) | 1,031,893 | 539,587 | POAHPS |
| | | | 471,198 | Benign |
| | | | 14,394 | DDoS |
| | | | 6,714 | C&C |

Before their data was usable, both datasets required a similar preprocessing stage. First, the features that did not provide any valuable information about a flow's benign or malicious purpose, such as timestamps and IP addresses, were discarded. Then, the categorical features were converted to numeric values by performing one-hot encoding. Due to the high cardinality of these features, the very low frequency categories were aggregated into a single category designated as 'Other', to avoid encoding qualitative values that were present in almost no samples and therefore had a small relevance.

Finally, the holdout method was applied to randomly split the data into training and evaluation sets with 70% and 30% of the samples, respectively. To ensure that the original class proportions were preserved, the split was performed with stratification. The resulting CIC-IDS2017 sets were comprised of 8 imbalanced classes and 83 features, 58 numerical and 25 categorical, whereas the IoT-23 sets contained 4 imbalanced classes and approximately half the structural size, with 42 features, 8 numerical and 34 categorical.

*4.2. Base Configurations*

After the data preprocessing stage, the distinct characteristics of the datasets were analyzed to identify the concrete constraints required for each scenario and establish the base configurations for A2PM.

Regarding CIC-IDS2017, some numerical features had discrete values that could only have integer perturbations. Due to the correlation between the encoded categorical features, they required combined perturbations to be compatible with a valid flow. Additionally, to guarantee the coherence of a generated flow with its type of cyber-attack, the encoded features representing the utilized communication protocol and endpoint, designated as port, could not be modified. Hence, the following configuration was used for the Enterprise scenario, after it was converted to the respective subset of feature indices:

1. Interval pattern – Modify {*numerical features*}, Integer {*discrete features*};
2. Combination pattern – Modify {*categorical features*}, Lock {*port, protocol*}.

Despite the different features of IoT-23, it presented similar constraints. The main difference was that, in addition to the communication protocol, a generated flow had to be coherent with the application protocol as well, which was designated as service. The base configuration utilized for the IoT scenario was:

1. Interval pattern – Modify {*numerical features*}, Integer {*discrete features*};
2. Combination pattern – Modify {*categorical features*}, Lock {*port, protocol, service*}.

It is pertinent to note that, for the 'Benign' class, A2PM would only generate benign network traffic that could be misclassified as a cyber-attack. Therefore, the configurations were only applied to the malicious classes, to generate examples compatible with their malicious purposes. Furthermore, since the examples should resemble the original flows as much as possible, the 'probability to be applied' was 0.6 and 0.4 for the interval and combination patterns, respectively. These values were established to slightly prioritize the small-scale modifications of individual numerical features over the more significant modifications of combined categorical features.

*4.3. Models and Fine-tuning*

A total of four MLP and four RF classifiers were created, one per scenario and training approach: regular or adversarial training. The first approach used the original training sets, whereas the latter augmented the data with one adversarial example per malicious flow. To prevent any bias, the examples were generated by adapting A2PM solely to the training data. The models and their fine-tuning process are described below.

An MLP [39] is a feedforward ANN consisting of an input layer, an output layer and one or more hidden layers in between. Each layer can contain multiple nodes with forward connections to the nodes of the next layer. When utilized as a classifier, the number of input and output nodes correspond to the number of features and classes, respectively, and a prediction is performed according to the activations of the output nodes.

Due to the high computational cost of training an MLP, it was fine-tuned using a Bayesian optimization technique [40]. A validation set was created with 20% of a training set, which corresponded to 14% of the original samples. Since an MLP accounts for the loss of the training data, the optimization sought to minimize the loss of the validation data. To prevent overfitting, early stopping was employed to end the training when this loss stabilized. Additionally, due to the class imbalance present in both datasets, the assigned class weights were inversely proportional to their frequency.

The fine-tuning led to a four-layered architecture with a decreasing number of nodes for both training approaches. The hidden layers relied on the computationally efficient Rectified Linear Unit (ReLU) activation function and the dropout technique, which inherently prevents overfitting by randomly ignoring a certain percentage of the nodes during training. To address multi-class classification, the Softmax activation function was used to normalize the outputs to a class probability distribution. The MLP architecture for the Enterprise scenario was:

1. Input layer – 83 nodes, 512 batch size;
2. Hidden layer – 64 nodes, ReLU activation, 10% dropout;
3. Hidden layer – 32 nodes, ReLU activation, 10% dropout;
4. Output layer – 8 nodes, Softmax activation.

A similar architecture was utilized for the IoT scenario, although it presented a decreased batch size and an increased dropout:

1. Input layer – 42 nodes, 128 batch size;
2. Hidden layer – 32 nodes, ReLU activation, 20% dropout;
3. Hidden layer – 16 nodes, ReLU activation, 20% dropout;
4. Output layer – 4 nodes, Softmax activation.

The remaining parameters were common to both scenarios because of their equivalent classification tasks. Table 3 summarizes the MLP configuration.

**Table 3.** Summary of Multilayer Perceptron configuration.

| Parameter | Value |
| --- | --- |
| Objective Loss | Categorical Cross-Entropy |
| Optimizer | Adam Algorithm |
| Learning Rate | 0.001 |
| Maximum Epochs | 50 |
| Class Weights | Balanced |

On the other hand, an RF [41] is an ensemble of decision trees, where each individual tree performs a prediction according to a different feature subset, and the most voted class is chosen. It is based on the wisdom of the crowd, the idea that a multitude of classifiers will collectively make better decisions than just one.

Since training an RF has a significantly lower computational cost, a five-fold cross-validated grid search was performed with well-established hyperparameter combinations. In this process, five stratified subsets were created, each with 20% of a training set. Then, five distinct iterations were performed, each training a model with four subsets and evaluating it with the remaining one. Hence, the MLP validation approach was replicated five times per combination. The macro-averaged F1-Score, which will be described in the next subsection, was selected as the metric to be maximized. Table 4 summarizes the optimized RF configuration, common to both scenarios and training approaches.

**Table 4.** Summary of Random Forest configuration.

| Parameter | Value |
| --- | --- |
| Splitting Criteria | Gini Impurity |
| Number of Trees | 100 |
| Maximum Depth of a Tree | 32 |
| Minimum Samples in a Leaf | 2 |
| Maximum Features | $\sqrt{\text{Number of Features}}$ |
| Class Weights | Balanced |

*4.4. Attacks and Evaluation Metrics*

A2PM was applied to perform adversarial attacks against the fine-tuned models for a maximum of 50 iterations, by adapting to the data of the holdout evaluation sets. The attacks were untargeted, causing any misclassification of malicious flows to different classes, as well as targeted, seeking to misclassify malicious flows as the 'Benign' class. To perform a trustworthy evaluation of the impact of the generated examples on a model's performance, it was essential to select appropriate metrics. The considered metrics and their interpretation are briefly described below [42], [43].

Accuracy measures the proportion of correctly classified samples. Even though it is the standard metric for classification tasks, its bias towards the majority classes must not be disregarded when the minority classes are particularly relevant to a classification task, which is the case of network-based intrusion detection [44]. For instance, in the Enterprise scenario, 77% of the samples have the 'Benign' class label. Since A2PM was configured to not generate examples for that class, even if an adversarial attack was successful and all generated flows evaded detection, an accuracy score as high as 77% could still be achieved. Therefore, to correctly exhibit the misclassifications caused by the performed attacks, the accuracy of a model was calculated using the network flows of all classes except 'Benign'. This metric can be expressed as:

$$Accuracy = \frac{TP + TN}{TP + TN + FP + FN} \tag{5}$$

where $TP$ and $TN$ are the number of true positives and negatives, correct classifications, and $FP$ and $FN$ are the number of false positives and negatives, misclassifications.

Despite the reliability of accuracy for targeted attacks, it does not entirely reflect the impact of the performed untargeted attacks. Due to their attempt to cause any misclassification, their impact across all the different classes must also be measured. The F1-Score calculates the harmonic mean of precision and recall, considering both false positives and false negatives. To account for class imbalance, it can be macro-averaged, which gives all classes the same relevance. This is a reliable evaluation metric because a score of 100% indicates that all cyber-attacks are being correctly detected and there are no false alarms. Additionally, due to the multiple imbalanced classes present in both datasets, it is also the most suitable validation metric for the employed fine-tuning approach. The macro-averaged F1-Score is mathematically defined as:

$$Macro\text{-}averaged\ F1\text{-}Score = \frac{1}{C} * \sum_{i=1}^{C} \frac{2 * P_i * R_i}{P_i + R_i} \tag{6}$$

where $P_i$ and $R_i$ are the precision and recall of class $i$, and $C$ is the number of classes.

*4.5. Enterprise Scenario Results*

In the Enterprise network scenario, adversarial cyber-attack examples were generated using the original flows of the CIC-IDS2017 dataset. The results obtained for the targeted and untargeted attacks were analyzed, and assessments of example realism and time consumption were performed. To assess the realism of the generated examples, these were analyzed and compared with the corresponding original flows, considering the intricacies and malicious purposes of the cyber-attacks. In addition to A2PM, the assessment included its potential alternatives: JSMA and OnePixel. To prevent any bias, a randomly generated number was used to select one example, detailed below.

The selected flow had the 'Slowloris' class label, corresponding to a Denial-of-Service attack that attempts to overwhelm a web server by opening multiple connections and maintaining them as long as possible [45]. The data perturbations created by A2PM increased the total flow duration and the packet Inter-Arrival Time (IAT), while reducing the number of packets transmitted per second and their size. These modifications were mostly focused on enhancing time-related aspects of the cyber-attack, to prevent its detection. Hence, in addition to being valid network traffic that can be transmitted through a computer network, the adversarial example also remained coherent with its class.

On the other hand, JSMA could not generate a realistic example for the selected flow. It created a major inconsistency in the encoded categorical features by assigning a single network flow to two distinct communication endpoints: destination ports number 80 (P80) and 88 (P88). Due to the unconstrained perturbations, the value of the feature representing P88 was increased without accounting for its correlation with P80, which led to an invalid example. In addition to the original Push flag (PSH) to keep the connection open, the method also assigned the Finished flag (FIN), which signals for connection termination and therefore contradicts the cyber-attack's purpose. Even though two numerical features were also slightly modified, the adversarial example could only evade detection by using categorical features incompatible with real network traffic.

Similarly, OnePixel also generated an example that contradicted the 'Slowloris' class. The feature selected to be perturbed represented the Reset flag (RST), which also causes termination. Since the method intended to perform solely one modification, it increased the value of a feature that no model learnt to detect because it is incoherent with that cyber-attack. Consequently, neither JSMA nor OnePixel are adequate alternatives to A2PM for tabular data. Table 5 provides an overview of the modified features. The '--' character indicates that the original value was not perturbed.

**Table 5.** Modified features of an adversarial 'Slowloris' example.

| Feature | Original Value | A2PM Value | JSMA Value | OnePixel Value |
|---|---|---|---|---|
| Flow duration | 109,034,141 | 119,046,064 | 109,034,140 | -- |
| Mean flow IAT | 13,600,000 | 19,374,259 | -- | -- |
| Flow packets per second | 0.0825 | 0.0429 | 0.0824 | -- |
| Mean forward packet length | 49.4 | 48.1 | -- | -- |
| Minimum forward segment size | 40 | 36 | -- | -- |
| Connection flags | 'PSH' | -- | 'PSH' + 'FIN' | 'PSH' + 'RST' |
| Destination port | 'P80' | -- | 'P80' + 'P88' | -- |

Regarding the targeted attacks performed by A2PM, the models created with regular training exhibited significant performance declines. Even though both MLP and RF achieved over 99% accuracy on the original evaluation set, a single iteration lowered their scores by approximately 15% and 33%. In the subsequent iterations, more malicious flows gradually evaded MLP detection, whereas RF was quickly exploited. After 50 iterations, their very low accuracy evidenced their inherent susceptibility to adversarial examples. In contrast, the models created with adversarial training kept significantly higher scores, with fewer flows being misclassified as benign. By training with one generated example per malicious flow, both classifiers successfully learned to detect most cyber-attack variations. RF stood out for preserving the 99.91% it obtained on the original data throughout the entire attack, which highlighted its excellent generalization (Figure 5).

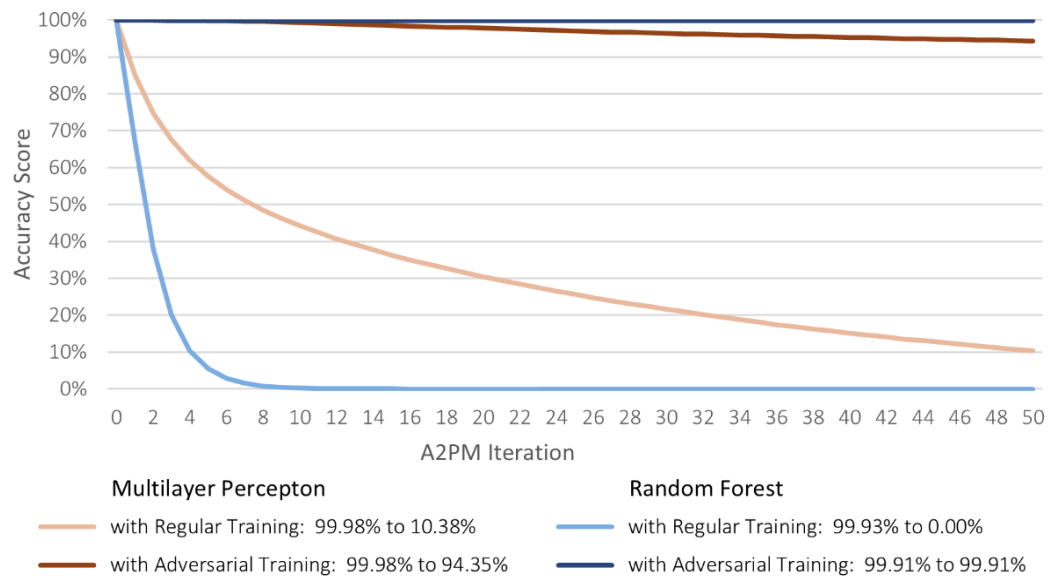

**Figure 5.** Targeted attack accuracy of Enterprise network scenario.

The untargeted attacks significantly lowered both evaluation metrics. The accuracy and macro-averaged F1-Score declines of the regularly trained models were approximately 99% and 79%, although RF was more affected in the initial iterations. The inability of both classifiers to distinguish between the different classes corroborated their high susceptibility to adversarial examples. Nonetheless, when adversarial training was performed, the models preserved considerably higher scores, with a gradual decrease of less than 2% per iteration. Despite some examples still deceiving them into predicting incorrect classes, both models were able to learn the intricacies of each type of cyber-attack, which mitigated the impact of the created data perturbations. The adversarially trained RF consistently reached higher scores than MLP in both targeted and untargeted attacks, indicating a better robustness (Figures 6 and 7).

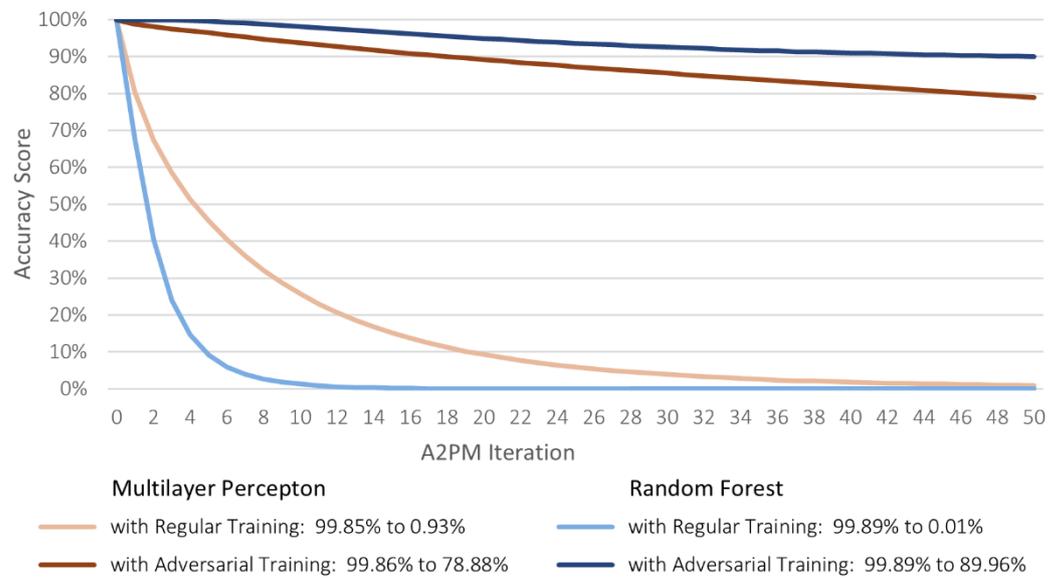

**Figure 6.** Untargeted attack accuracy of Enterprise network scenario.

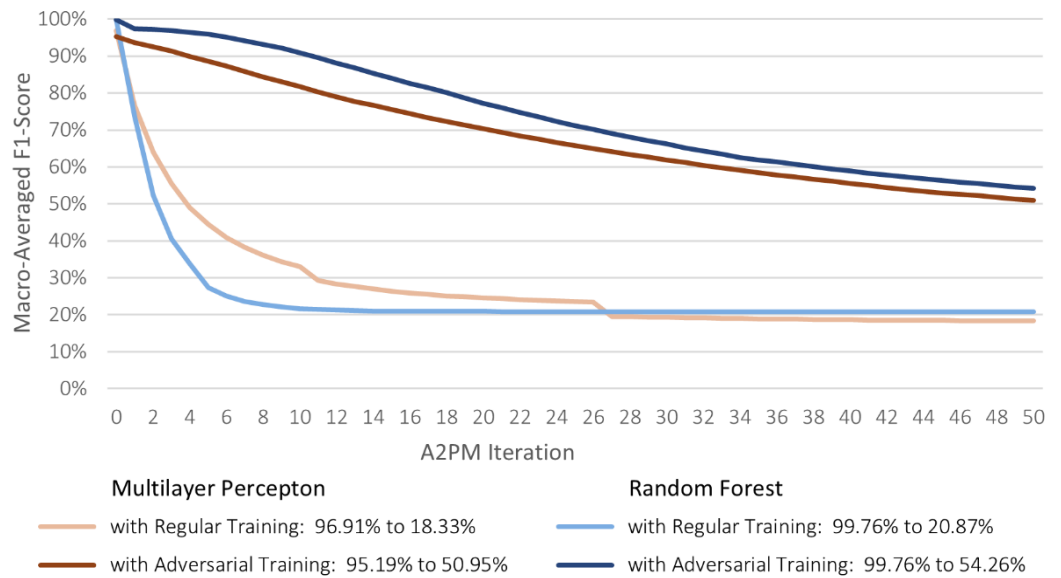

**Figure 7.** Untargeted attack F1-Score of Enterprise network scenario.

To analyze the time consumption of A2PM, the number of milliseconds required for each iteration was recorded and averaged, accounting for the decreasing quantity of new examples generated as an attack progressed. The generation was performed at rate of 10 examples per 1.7 milliseconds on the utilized hardware, which evidenced the fast execution and scalability of the proposed method when applied to adversarial training and attacks in Enterprise computer networks.

*4.6. IoT Scenario Results*

In the IoT network scenario, the adversarial cyber-attack examples were generated using the original flows of the IoT-23 dataset. The analysis performed for the previous scenario was replicated to provide similar assessments, including the potential alternatives of the current literature: JSMA and OnePixel.

The randomly selected flow for the assessment of example realism had the 'DDoS' class label, which corresponds to a Distributed Denial-of-Service attack performed by the malwares recorded in the IoT-23 dataset. A2PM replaced the encoded categorical features of the connection state and history with another valid combination, already used by other original flows of the 'DDoS' class. Instead of an incomplete connection (OTH) with a bad packet checksum (BC), it became a connection attempt (S0) with a Synchronization flag (SYN). Hence, the generated network flow example remained valid and compatible with its intended malicious purpose, achieving realism.

As in the previous scenario, both JSMA and OnePixel generated unrealistic examples. Besides the original OTH, both methods also increased the value of the feature representing an established connection with a termination attempt (S3). Since a flow with simultaneous OTH and S3 states is neither valid nor coherent with the cyber-attack's purpose, the methods remain inadequate alternatives to A2PM for tabular data. In addition to the states, JSMA also assigned a single flow to two distinct communication protocols, Transmission Control Protocol (TCP) and Internet Control Message Protocol (ICMP), which further evidenced the inconsistency of the created data perturbations. Table 6 provides an overview of the modified features, with '--' indicating an unperturbed value.

**Table 6.** Modified features of an adversarial 'DDoS' example.

| Feature | Original Value | A2PM Value | JSMA Value | OnePixel Value |
|---|---|---|---|---|
| Connection state | 'OTH' | 'S0' | 'OTH' + 'S3' | 'OTH' + 'S3' |
| Connection history | 'BC' | 'SYN' | -- | -- |
| Communication protocol | 'TCP' | -- | 'TCP' + 'ICMP' | -- |

Regarding the targeted attacks, A2PM caused much slower declines than in the previous scenario. The accuracy of the regularly trained MLP only started being lower than 50% at iteration 43, and RF stabilized with approximately 86%. These scores evidenced the decreased susceptibility of both classifiers, especially RF, to adversarial examples targeting the 'Benign' class. Furthermore, with adversarial training, the models were able to preserve even higher rates during an attack. Even though many examples still evaded MLP detection, the number of malicious flows predicted to be benign by RF was significantly lowered, which enabled it to keep its accuracy above 99%. Hence, the latter successful detected most cyber-attack variations (Figure 8).

The untargeted attacks iteratively caused small decreases of both metrics. Despite RF starting to stabilize from the fifth iteration forward, MLP continued its decline for an additional 48% of accuracy and 17% of macro-averaged F1-Score. This difference in both targeted and untargeted attacks suggests that RF, and possibly tree-based algorithms in general, have a better inherent robustness to adversarial examples of IoT network traffic. Unlike in the previous scenario, adversarial training did not provide considerable improvements. Nonetheless, the augmented training data still contributed to the creation of more adversarially robust models because they exhibited fewer incorrect class predictions throughout the attack (Figures 9 and 10).

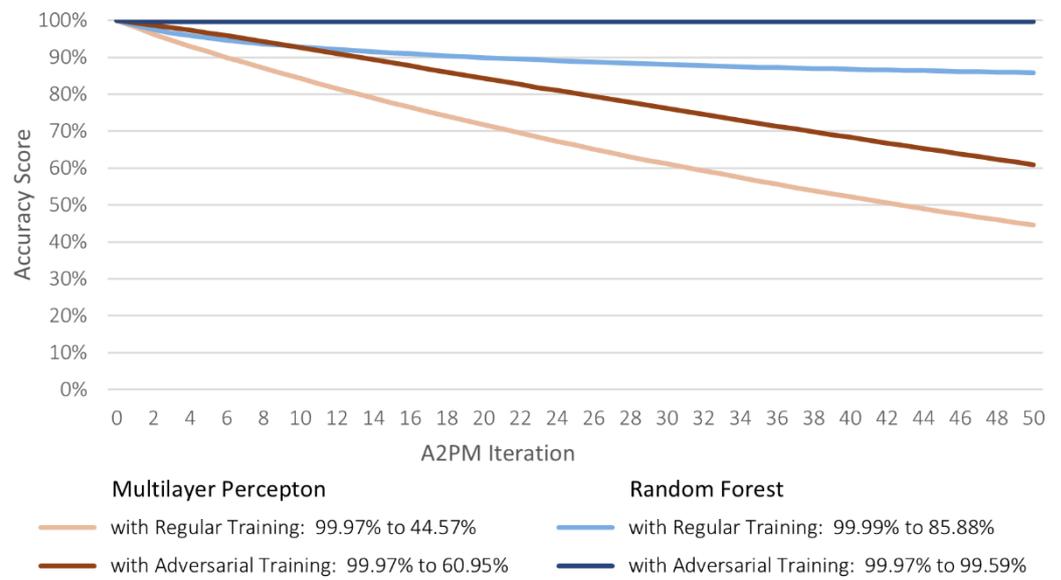

**Figure 8.** Targeted attack accuracy of IoT network scenario.

A time consumption analysis was also performed, to further analyze the scalability of A2PM on relatively common hardware. The number of milliseconds required for each iteration was recorded and averaged, resulting in a rate of 10 examples per 2.4 milliseconds. By comparing the rate obtained in both scenarios, it can be observed that it was 41% higher for IoT-23 than for CIC-IDS2017. Even though the former dataset had approximately half the structural size, a greater number of locked categorical features were provided to the Combination pattern. Therefore, the increased rate suggests that the more complex inter-feature constraints are specified, the more time will be required to apply A2PM. Nonetheless, the time consumption was still reasonably low, which further evidenced the fast execution and scalability of the proposed method.

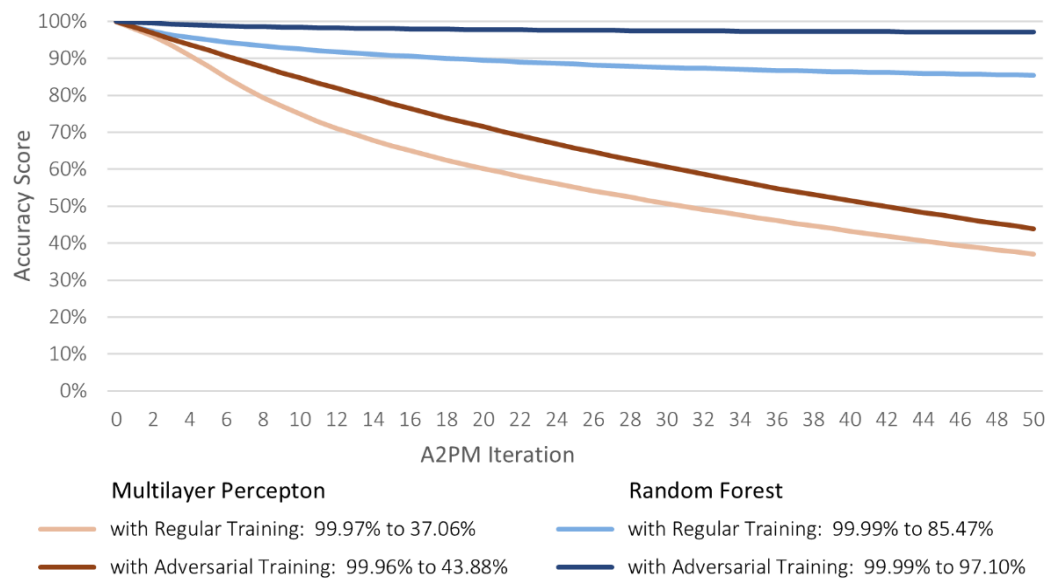

**Figure 9.** Untargeted attack accuracy of IoT network scenario.

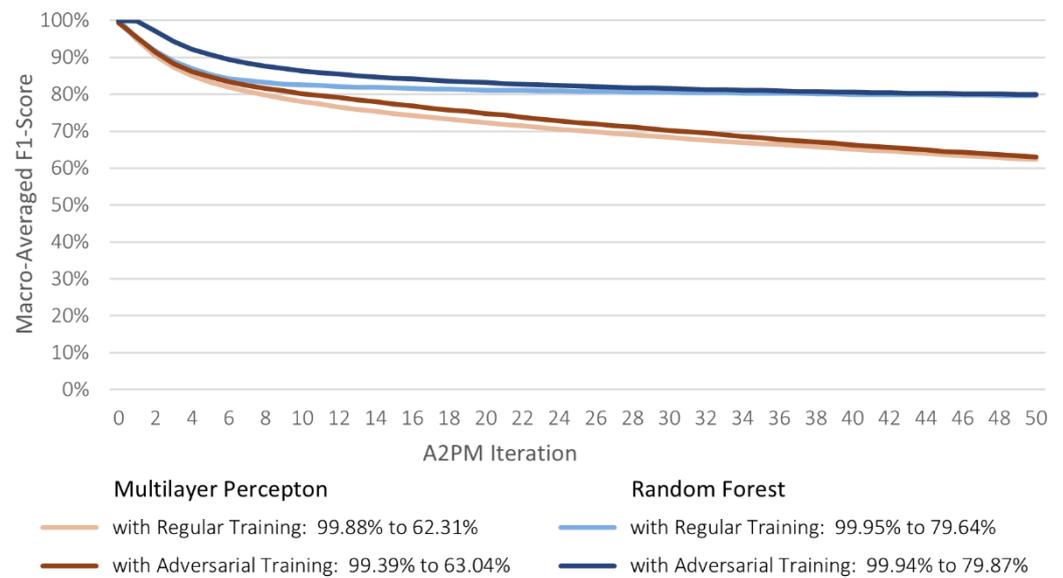

**Figure 10.** Untargeted attack F1-Score of IoT network scenario.

## 5. Conclusions

This work established the domain and class-specific constraint levels, which an adversarial example must comply with to achieve realism on tabular data, and introduced A2PM to fulfil these constraints in a gray-box setting, with only knowledge of the feature set. The capabilities of the proposed method were evaluated in a cybersecurity case study with two scenarios: Enterprise and IoT networks. MLP and RF classifiers were created with regular and adversarial training, using the network flows of the CIC-IDS2017 and IoT-23 datasets, and targeted and untargeted attacks were performed against them. For each scenario, the impact of the attacks was analyzed, and assessments of example realism and time consumption were performed.

The modular architecture of A2PM enabled the creation of pattern sequences adapted to each type of cyber-attack, according to the concrete constraints of the utilized datasets. Both targeted and untargeted attacks successfully decreased the performance of all MLP and RF models, with significantly higher declines exhibited in the Enterprise scenario. Nonetheless, the inherent susceptibility of these models to adversarial examples was mitigated by augmenting their training data with one generated example per malicious flow. Overall, the obtained results demonstrate that A2PM provides a scalable generation of valid and coherent examples for network-based intrusion detection. Therefore, the proposed method can be advantageous for adversarial attacks, to iteratively cause misclassifications, and adversarial training, to increase the robustness of a model.

In the future, the patterns can be improved to enable the configuration of more complex intra and inter-feature constraints. Since it is currently necessary to use both Interval and Combination patterns to perturb correlated numerical features, a new pattern can be developed to address their required constraints. It is also imperative to analyze other datasets and other domains to contribute to robustness research. Future case studies can further reduce the knowledge required to create realistic examples.

**Author Contributions:** Conceptualization, J.V., N.O. and I.P.; methodology, J.V. and N.O.; software, J.V.; validation, N.O. and I.P.; investigation, J.V. and I.P.; writing, J.V. and I.P.; supervision, I.P.; project administration, I.P.; funding acquisition, I.P. All authors have read and agreed to the published version of the manuscript.

**Funding:** The present work has received funding from the European Union's Horizon 2020 research and innovation program, under project SeCoIIA (grant agreement no. 871967). This work has also received funding from UIDP/00760/2020.

**Data Availability Statement:** Publicly available datasets were analyzed in this work. The data can be found at: CIC-IDS2017, IoT-23. A novel method was developed in this work. An implementation in the Python 3 programming language can be found at: A2PM.

**Conflicts of Interest:** The authors declare no conflict of interest. The funders had no role in the design of the study; in the collection, analyses, or interpretation of data; in the writing of the manuscript, or in the decision to publish the results.

*Future Internet Preprint* 18 of 18